4Dia: A tool for automated 4D microscope image alignment

Nimmy S. John[1], Michelle A. Urman[1] and ChangHwan Lee[1,2]

[1] Department of Biological Sciences, the RNA Institute, University at Albany, New York, NY, USA

[2] Corresponding author

Highlights

- 4Dia aligns three-dimensional time-lapse images both in Z-stacks and over time.
- 4Dia can use any channel for image alignment and can be applied to images of any tissue sample.
- 4Dia is an ImageJ/Fiji macro script that can be easily customized.

Abstract

Recent advances in microscopy enable three-dimensional live imaging at a high resolution. Long-term live imaging of a multicellular organism requires immobilization of the organism under stable physiological conditions. Despite proper immobilization, challenges remain within live imaging data analysis due to other intrinsic and extrinsic dynamics, which can result in misalignments of an image series over time. 4Dia, an ImageJ/Fiji macro script, aligns 3D timelapse images through Z-stacks as well as over time using any user selected channel. 4Dia can be used for essentially any tissue sample with no limit on the size of Z-stack or the number of timepoints.

| Code metadata | |
| --- | --- |
| Current code version | v1.1 |
| Permanent link to code/repository used for this code version | |
| Permanent link to Reproducible Capsule | |
| Legal Code License | MIT License |
| Code versioning system used | Git |
| Software code languages, tools, and services used | IJM |
| Compilation requirements, operating environments & dependencies | Fiji (ImageJ v1.53f), 'MultiStackReg' plugin, and 'TurboReg' plugin |
| If available Link to developer documentation/manual | https://github.com/chleelab/4Dia |
| Support email for questions | chlee@albany.edu |

1. Introduction

Since the advent of optical microscopy, imaging has become a crucial tool that is used daily in various subfields of biology, including cell biology and genetics. Particularly, fluorescence microscopy, which uses fluorescent proteins or dyes visible at specific wavelengths, provides highly sensitive, specific, and reliable microscope images that enable observation of organelles and biomacromolecules of interest with great detail[1,2]. The recent advances in optical techniques have brought fluorescence microscopy to another level, where macromolecules, such as proteins and nucleic acids, can be visualized at a single-molecule resolution[3]. These breakthroughs in microscopy and the advances in mechanics have led to high-resolution imaging both in space and time[4–10]. A microscope with a fast, sensitive camera or photodetector and a precise motorized stage can now achieve sub-molecular resolution not just in space with a finer pixel/voxel size or a higher signal-to-noise ratio but also in temporal scale. High spatial resolution typically requires extended imaging time (e.g., long laser dwelling time for confocal laser scanner) and thus is the most beneficial to Imaging with fixed samples, where the temporal factor can be ignored. In contrast, high temporal resolution is crucial for live imaging, especially for capturing highly dynamic events like transcriptional bursting in real time or working with an organism that moves rapidly.

Live imaging of intact organisms has been considerably more challenging than fixed sample imaging. Extrinsic factors such as any stage and focus drift, mobility, and stable physiological conditions (temperature, mechanical pressure, etc.) must be considered when performing long-term live imaging of intact multicellular organisms. Several live imaging methods have been developed with these factors in consideration, but little was applicable to multicellular organisms that are mobile until recently[11,12]. However, even if the organisms are properly immobilized and well maintained within the imaging system, cells and tissues can still move within the living system. For example, some cells migrate toward another region of the tissue, or a tissue of interest can be shifted by an adjacent tissue that has some dynamics such as intestinal movements[11]. All these factors make it even more difficult to monitor dynamic biological phenomena for a long period of time. However, these intrinsic movements cannot be easily regulated, as disruption of such activities often creates artifacts. Therefore, computational processing after image acquisition to minimize the effects of these movements is a critical step to obtain a precise and complete analysis of live imaging data.

Here, we describe an ImageJ macro for automated 4D microscope image alignment (4Dia), which we developed to align the tissue or organism through space (in X-, Y-, and Z-axis) and time. The macros were created using images of the *Caenorhabditis elegans* germline but can be applied to essentially any tissue type[11].

2. Description

The 4D microscope image alignment (4Dia) uses three-dimensional Z-stack timelapse images (TIFF format) and aligns them along the Z-axis, as well as through time using any channel as a reference for alignment, including a fluorescent channel (e.g., GFP), brightfield, or DIC channel. There is no limit to the number of slices in the Z-stack, channels, or time points that 4Dia can handle. 4Dia works within the ImageJ/Fiji software and requires two ImageJ plugins, 'MultiStackReg' (http://bradbusse.net/downloads.html) and 'TurboReg' (http://bigwww.epfl.ch/thevenaz/turboreg/), which must be pre-installed to run 4Dia.

4Dia comprises two ImageJ/Fiji macros: (1) 'MultiStackReg_BatchProcess1of2.ijm' and (2) 'MultiStackReg_BatchProcess2of2.ijm', which must be run sequentially. The first macro aligns the Z-stack images at each timepoint through the Z-axis. This process is particularly useful for highly dynamic tissues or organisms even when an image stack is acquired within one timepoint. The second macro uses the processed images from the first macro and further aligns the Z-stacks through all timepoints.

(1) 'MultiStackReg_BatchProcess1of2.ijm'

Before running the macro, the user needs to specify the channel that will be used for image alignment. To do so, open the file using ImageJ/Fiji or a text editor and assign the number that matches the order of the channels to the variable "ChToUse" in line 7. For example, if the second channel will be used for alignment, then the user should assign the number "2" to the variable ("ChToUse = 2"). When the macro is executed, the user is first asked to choose a source directory as well as an output folder on a pop-up window. Once these two folders are selected, the macro then cycles through the timelapse Z-stack images and processes them one by one. Any files in a format other than TIFF in the source folder will be excluded from alignment. The macro separates out the channel to be used for alignment and start aligning the Z-slices along the Z-stack using the similarity in the signal pattern between two adjacent Z-slices. The alignment begins from the middle of the Z-stack, using it as a reference image. The macro applies the same alignment vectors to all other channels, combines channels, and saves the new aligned image file with the original file name with a '_Zreg' suffix in the designated output folder. If the timelapse Z-stack images do not require alignment through the Z-axis, this first macro can be skipped.

(2) 'MultiStackReg_BatchProcess2of2.ijm'

This macro works similarly to the first macro described above. The user has to specify the channel that will be used for image alignment, as explained above for the first macro. Upon execution, the macro opens a dialog box to input the source directory and the output folder. The newly generated folder by the first macro should be chosen to further align the images through timepoints. This macro uses the maximum Z-projection to align images between timepoints. It separates each channel, records the alignment vectors for each timepoint using Z-projected images from the first channel (per current settings), and applies them to all Z-slices on each timepoint. The new aligned images are saved with a '_Treg' suffix in the designated output folder.

3. Impact

Long-term live imaging of intact organisms remains difficult to analyze due to intrinsic and extrinsic movement. Live imaging data, especially long-term 3D timelapse images, are often very large as each timepoint contains a whole Z-stack. In addition, the quantity of data to analyze quickly expands when other conditions such as mutants are added to the live imaging experiment. Therefore, an automated, systematic tool is required to ensure the live imaging data are ready to be analyzed. One of the most critical steps before image analysis is to align all Z-stacks through timelapse to easily trace dynamics of biological phenomena such as transcriptional bursting events[11,12]. To our knowledge, 4Dia is the first script that aligns the 3D Z-stack images both through the Z-axis as well as timepoints in an automated fashion. 4Dia allows 4D image alignment through the Z-stack and timepoints using any channel of choice such as a fluorescent, brightfield, or DIC channel. 4Dia was initially developed to align *C. elegans* germline tissue images to monitor transcriptional dynamics[11]. However, 4Dia can be used for essentially any tissue type with no limit with the size of the Z-stack or the number of timepoints. 4Dia can help researchers in any field that use a microscope or biomedical imaging and need to process their timelapse images before analysis for a more precise assessment of dynamics.

4. Limitations and potential improvements

The main limitation of the scripts is that they have not been tested for images other than the *C. elegans* germline. However, we expect that the macros would work for all tissue types as they are not based on any specific features of the *C. elegans* germline tissue. One of our goals is to use other images with different species and signals to further optimize the scripts. Another potential limitation is that the scripts might not be able to align tissues that undergo dramatic movement. We implemented another macro to manually fix such an issue by transforming a specific Z-slice or timepoints by a user input, which is available at: https://github.com/chleelab/4Dia. The latest versions of 4Dia will also be uploaded when any updates are available.

Declaration of Competing Interest

The authors declare that they have no known competing financial interests or personal relationships that could have appeared to influence the work reported in this paper.

Acknowledgments

We thank Judith Kimble, Sarah L. Crittenden, and Tina R. Lynch for their comments on this manuscript. This project was supported by American Heart Association (18POST34030263, CHL) and the RNA Institute Fellowship (MU).